\documentclass[12pt]{article}
\usepackage{graphicx}

\newcommand{\beq}{\begin{equation}}
\newcommand{\eeq}{\end{equation}}
\newcommand{\beqn}{\begin{eqnarray}}
\newcommand{\eeqn}{\end{eqnarray}}
\newcommand{\beqns}{\begin{eqnarray*}}
\newcommand{\eeqns}{\end{eqnarray*}}

\begin{document}
\begin{titlepage}
\begin{center}

\hfill USTC-ICTS/PCFT-21-09\\
\hfill February 2021\\

\vspace{2.5cm}

{\large {\bf  Exclusive rare Higgs decays into lepton pair and light mesons}} \vspace*{1.0cm}\\
{  Dao-Neng Gao$^\dagger$ and Xi Gong$^\ddagger$} \vspace*{0.3cm} \\
{\it\small
Interdisciplinary Center for Theoretical Study, University of Science and Technology of China,
Hefei, Anhui 230026 China}\\
{\it\small Peng Huanwu Center for Fundamental Theory, Hefei, Anhui 230026 China}
\vspace*{1cm}
\end{center}
\begin{abstract}
\noindent
Exclusive rare Higgs decays into lepton pair plus one light hadron, such as $h\to \rho^0(\omega)\ell\bar{\ell}$, $h\to \pi^0\ell\bar{\ell}$, $h\to \pi^+ (K^+)\ell^-\bar{\nu}_\ell$, and $h\to \rho^+(K^{*+})\ell^-\bar{\nu}_\ell$, have been explored in the standard model. Decay amplitudes are dominantly from the Higgs couplings to gauge bosons and to charged leptons, and their branching ratios are predicted in the range of $10^{-8}\sim 10^{-5}$. We have also analyzed the differential dilepton invariant mass and angular distributions of $h\to \rho^0 \ell^+\ell^-$ decays. It will be challenging to search for these rare processes. Nevertheless, experimental studies of them, in particular, $h\to \rho^0\ell^+\ell^-$ with $\ell=e,\mu$, might be interesting both to help deepen our understanding of the standard model and to probe new physics beyond the standard model in the future high-precision experiments.
\end{abstract}

\vfill

\noindent
$^{\dagger}$ Email address: ~gaodn@ustc.edu.cn\\\noindent
$^{\ddagger}$ Email address: ~gonff@mail.ustc.edu.cn

\end{titlepage}

\section{Introduction}

The discovery of the 125 GeV Higgs boson by the ATLAS and CMS collaborations \cite{atlascms} at the CERN Large Hadron Collider (LHC) in 2012 was a major breakthrough in particle physics, which completes the standard model (SM) and opens up a new era of the precise determination of the properties of this new particle as well. So far, experimental studies of the Higgs boson couplings to the SM fields \cite{atlascmsjoint, cmsatlas19} show no significant deviations from the SM predictions. Nevertheless, it is conceivable that much more detailed investigations both theoretically and experimentally, may help to reveal the non-standard properties of the particle, which would be very useful to increase our understanding of Higgs dynamics.

Decays of the Higgs boson into gauge bosons including $h\to \gamma\gamma$, $h\to Z Z^*$, and $h\to W W^*$, play important roles in the discovery of the particle. In addition to improving the measurements of these modes, exclusive rare Higgs decays would be also very interesting at the future high energy experimental facilities, such as the high-luminosity LHC and high-energy LHC, Higgs factory, even 100 TeV proton-proton collider, in which one could have a large sample of the Higgs particle. Actually, some types of these decays have been studied theoretically and experimentally, like $h\to V \gamma$ decays \cite{WK7783,BPSV13, KPPSSZ14, KN15, atlas15} with  $V$ denoting vector mesons $\rho$, $\phi$, $J/\psi$ etc., and $h\to V Z$ decays \cite{IMT13, Gao,BAL14,MS14,AKN16,Zhao18}, as well as leptonic final states processes $h\to\gamma \ell^+\ell^-$ decays \cite{ABDR97,FS07, DR13-14,CQZ13,KK13,SCG13,Pass13,KK14,CFLV14,HW17,KUN20, cms1316atlas14}, $h\to V \ell \bar{\ell}$ decays ($V=\Upsilon, J/\psi, \phi$) \cite{CFS16,PS16}, and $h\to \eta_{c,b} \ell^+\ell^-$ decays \cite{BMPS17}. In the SM these transitions have small branching fractions, the experimental study of them is generally a difficult task. Searches for these rare processes however may potentially probe the novel Higgs couplings in the case that their decay rates could be enhanced in some new scenarios beyond the SM.

\begin{figure}[t]
\begin{center}
\includegraphics[width=12cm,height=5cm]{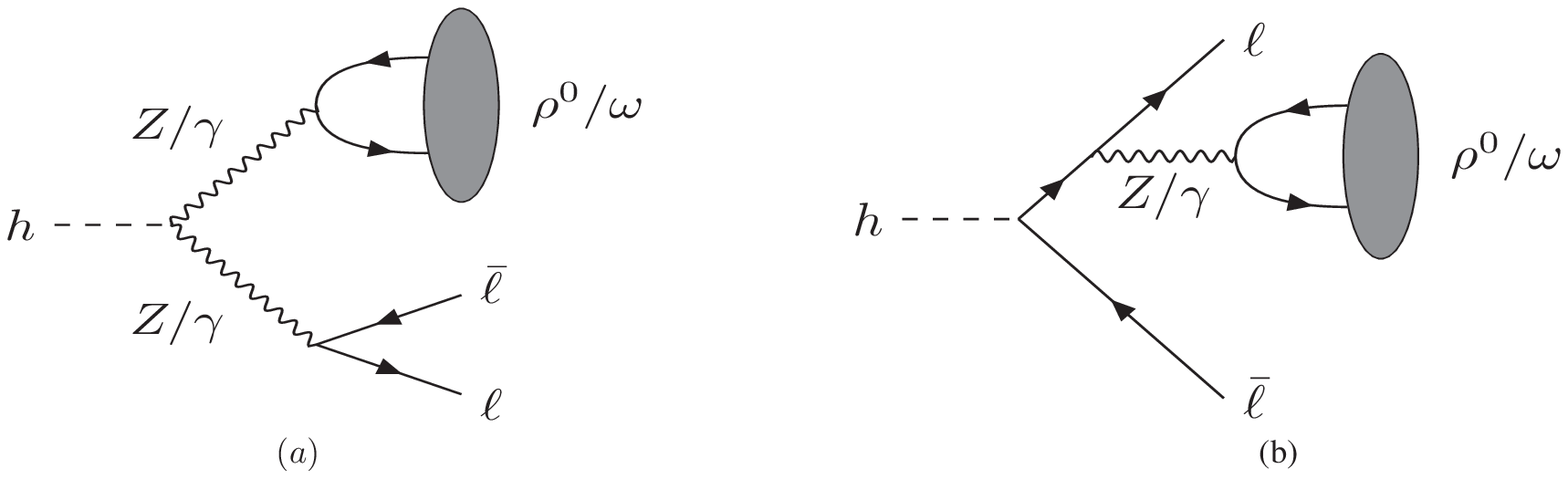}
\end{center}
\caption{Lowest-order diagrams for $h\to V\ell\bar{\ell}$ ($V=\rho^0, \omega$) decays in the SM.
 The virtual photon $\gamma$ or $Z$ can also be emitted from $\bar{\ell}$ line in the diagram (b). Only the diagram (a) involving $hZZ$ and $hZ\gamma$ vertices can contribute to $h\to V\bar{\nu}\nu$ transitions. % To get the leading-order contribution to the $h\to\pi^0\ell\bar{\ell}$ decays, we need only the diagram (a) involving $hZZ$ vertex.
 }\label{figure1}
\end{figure}

In the present paper we will focus on rare Higgs decays into lepton pair plus one light hadron containing $\rho$, $\omega$, and $\pi$ mesons. In the SM, these processes, such as $h\to V\ell\bar{\ell}$ ($V=\rho^0, \omega$) decays,  will get contribution from the diagrams, as shown in Figure \ref{figure1},  in which the Higgs boson couples to fermions and to the gauge bosons including $Z$ and $\gamma$. Interaction vertices of Higgs couplings to the SM fermions will be strongly suppressed for the electron and light quarks ($u$ and $d$) since they are proportional to $m_f/v$ ($v=246$ GeV is the vacuum expectation value of the Higgs field). For this reason we do not consider contributions generated from the Higgs coupling to $u$,$d$ quarks, and the second diagram of Figure \ref{figure1} can be neglected for the electron mode.

Higgs decays into a heavy meson plus lepton pair have been analyzed in Refs. \cite{CFS16, PS16, BMPS17}, and in the SM these branching ratios have been given as $10^{-6}\sim 10^{-8}$. It will be shown below that, for the light meson final states, their branching ratios are also around this range, the present work thus provides some complementary information for these previous studies. On the other hand, experimentally heavy quarkonia will be in general reconstructed via leptonic decays into muon pairs with relative small rates: ${\rm Br}(J/\psi\to \mu^+\mu^-)=(5.961\pm0.033)\%$ and ${\rm Br}(\Upsilon\to \mu^+\mu^-)=(2.48\pm0.05)\%$ \cite{PDG}; while for light mesons, $\rho^0$ decays almost exclusively to $\pi^+\pi^-$ and $\omega$ has a large rate into $\pi^+\pi^-\pi^0$ in the event reconstruction. Furthermore, since the contribution by the Higgs coupling to light quarks is negligible, most nonperturbative effects in our calculation, depicted in Figure \ref{figure1}(a) and (b), are confined to the matrix element
\beq\label{fv}
\langle V(p,\epsilon)|\bar{q}\gamma_\mu q |0\rangle= f_{V}m_{V}\epsilon^{*}_\mu,
\eeq
where $\epsilon^{*}_\mu$ is polarization vector of the vector meson $V$. $f_V$ is its decay constant, which can be extracted from the measured $V\to e^+e^-$ width. However, this is not the case for heavy meson final states, in which more nonperturbative information, such as the light-cone distribution amplitudes of $J/\psi$ and $\Upsilon$, would have to be involved in order to evaluate the contribution generated from  Figure 1(a) of Ref. \cite{CFS16}, due to the Higgs couplings to heavy quarks.

This paper is organized as follows. In the next section, we present a detailed derivation of decay amplitudes. Section 3 is our numerical analysis, including calculations of branching ratios and studies of differential decay rates.  We summarize our results and give some outlooks in Section 4.

\section{Decay amplitudes}

It is easy to see that, the diagram in Figure \ref{figure1}(a) contains the couplings of the Higgs boson to a pair of neutral gauge bosons $ZZ$, $Z\gamma$, and $\gamma\gamma$, which in turn are converted to a lepton pair and to a $q\bar{q}$ pair via the neutral current interactions
\beq\label{NC}
{\cal L}_{\rm NC}=e J_\mu^{\rm em} A^\mu+\frac{g}{\cos\theta_W}J^Z_\mu Z^\mu\eeq
with
\beq\label{emcurrent}
J_\mu^{\rm em}=\sum_f Q_f\bar{\psi}_f\gamma_\mu \psi_f,
\eeq
and
\beq\label{weakneutralcurrent}
J_\mu^Z=\frac{1}{2}\sum_f \bar{\psi}_f\gamma_\mu(g_V^f-g_A^f\gamma_5)\psi_f.\eeq
Here $e$ is the QED coupling constant, $g$ is the SU(2)$_L$ coupling constant, $\theta_W$ is the Weinberg angle, and $f$ denotes fermions including leptons and quarks.  Also $g_V^f=T_3^f-2 Q_f \sin^2 \theta_W$ and $g_A^f=T_3^f$,  where $Q_f$ is the charge, and $T_3^f$ is the third component of the weak isospin of the fermion. The $q\bar{q}$ pair then hadronizes into the vector meson $\rho$ or $\omega$.

In the SM, $Z$-boson can couple to the Higgs boson at the tree-level, and the $hZZ$ vertex is written as
\beq\label{hzz}{\cal L}_{h ZZ}=\frac{m_Z^2}{v}h Z_\mu Z^\mu.
\eeq
However, the leading-order SM $hZ\gamma$ and $h\gamma\gamma$ interactions are induced by one-loop diagrams involving $W$-boson or heavy charged fermions like top-quark, and their explicit expressions can be found in Ref. \cite{BH85}. On the other hand, one can write down the effective lagrangian for the $h Z \gamma$ and $h\gamma\gamma$ couplings generated in the SM as follows \cite{GHKD90}
\beq\label{effectivehzgamma}
{\cal L}^{hZ \gamma}_{\rm eff}=\frac{e g}{16\pi^2 v} C_{Z\gamma} Z_{\mu \nu} F^{\mu\nu} h,\eeq
and
\beq\label{effectivehgammagamma}
{\cal L}^{h\gamma \gamma}_{\rm eff}=\frac{e g}{32\pi^2 v} C_{\gamma\gamma} F_{\mu \nu} F^{\mu\nu} h,\eeq
where dimensionless coefficients $C_{Z\gamma}$ and $C_{\gamma\gamma}$ can be thought of as the effective couplings. Of course, for the general effective interactions beyond the SM, some new structures other than eqs. (\ref{effectivehzgamma}) and (\ref{effectivehgammagamma}) will appear, which have been analyzed in Refs. \cite{KK13, KN15, AKN16}.

\begin{table}[t]\begin{center}\begin{tabular}{ c c  c  c} \hline\hline
 Meson  & $G_{V}$& $Q_{V}$ & $f_{V/P}$[MeV]\\\hline
 $\rho^0$ & $\frac{1}{\sqrt{2}}\left(\frac{1}{2}-\sin^2\theta_W\right)$& $\frac{1}{\sqrt{2}}$ & $216.3\pm 1.3$
 \\\\
$\omega$& $\frac{-\sin^2\theta_W}{3\sqrt{2}}$ & $\frac{1}{3\sqrt{2}}$ &$194.2\pm 2.1$ \\
$\rho^\pm$ & & & $207.8\pm 1.4$\\
$K^{*\pm}$ & & & $203.2\pm 5.9$\\
$\pi$ & & & $130.4\pm 0.2$\\
$K^\pm$ & & & $156.2\pm 0.7$\\
\hline\hline
\end{tabular}\caption{Hadronic input parameters for light mesons. The values of $f_{V/P}$'s are taken from Ref. \cite{KN15}.} \end{center}\end{table}

Now let us go into the decay amplitudes. For the $h\to V \ell^+\ell^-$ transitions, direct calculation from Figure \ref{figure1} will give
\beqn\label{amplitude1}
{\cal M}(h\to V \ell^+\ell^-)&=&\frac{m_Z^2 g^2 G_V f_V m_V}{v \cos^2\theta_W(m_Z^2-m_V^2)}P_Z\epsilon^{*\mu}_V(p)\bar{u}_\ell(k_1)\gamma_\mu(g_V^\ell-g_A^\ell \gamma_5)v_\ell(k_2)\nonumber\\
&&-\frac{e^2g^2 C_{Z\gamma}Q_V f_V}{16\pi^2 v \cos\theta_W m_V}P_Z(p_\mu q_\nu-p\cdot q g_{\mu\nu})\epsilon^{*\nu}_V(p)\bar{u}_\ell(k_1)\gamma^\mu(g_V^\ell-g_A^\ell \gamma_5)v_\ell(k_2)\nonumber\\
&&+\frac{e^2g^2 C_{Z\gamma}G_V}{8\pi^2v\cos\theta_W(m_Z^2-m_V^2)}\frac{1}{q^2} (p_\mu q_\nu-p\cdot q g_{\mu\nu})\epsilon^{*\nu}_V(p)\bar{u}_\ell(k_1)\gamma^\mu v_\ell(k_2)\nonumber\\
&&-\frac{e^4 C_{\gamma\gamma}Q_V f_V}{8\pi^2v m_V}\frac{1}{q^2} (p_\mu q_\nu-p\cdot q g_{\mu\nu})\epsilon^{*\nu}_V(p)\bar{u}_\ell(k_1)\gamma^\mu v_\ell(k_2)\nonumber\\
&&-\frac{e^2 m_\ell Q_V f_V}{v m_V}\epsilon^{*\mu}_V\bar{u}_\ell(k_1)\left(\frac{2k_{1\mu}+\gamma_\mu p\!\!\!/}{2k_1\cdot p+m_\rho^2}-\frac{2k_{2\mu}+p\!\!\!/\gamma_\mu}{2k_2\cdot p+m_\rho^2}\right)v_\ell(k_2)
\eeqn
with
\beq\label{PZ}
P_Z=\frac{1}{m_Z^2-q^2-im_Z\Gamma_Z}
\eeq
parameterizing the $Z$ pole effect, where $k_1$, $k_2$ and $p$ represent the momentum of $\ell^-$, $\ell^+$ and $\rho$ in the final states, respectively. $q^2=(k_1+k_2)^2$ denotes the lepton pair mass squared, and $G_V$ and $Q_V$ are listed in Table 1. Eq. (\ref{fv}) and the vertex of the Higgs coupling to charged leptons
\beq\label{higgsleptons}
{\cal L}_{h\ell\bar{\ell}}=-\frac{m_\ell}{v}h\bar{\psi}_\ell\psi_\ell\eeq
have been used in deriving eq. (\ref{amplitude1}). It is obvious that the virtual $Z$ contribution from the diagram (b) of Figure \ref{figure1} is strongly suppressed, which has been neglected in eq. (\ref{amplitude1}). Similarly, for $h\to V \bar{\nu}\nu$ decays, only diagram (a) containing $hZZ$ and $hZ\gamma$ vertices can contribute, and the amplitudes can be read from the first two lines in eq. (\ref{amplitude1}). Note that $g_V^\nu=g_A^\nu=1/2$ for neutrino final states.

By squaring the amplitude and summing over the polarizations of final particles, the differential decay rate of $h\to V \ell^+\ell^-$ can be obtained as
\beq\label{distribution1}
\frac{d\Gamma}{d s d\cos\theta^{(\ell)}}=\frac{m_h}{512\pi^3}\beta_\ell \lambda^{1/2}(1, r_V^2,s) \sum_{\rm spins}|{\cal M}(h\to V\ell^+\ell^-)|^2
\eeq
 with $\beta_\ell=\sqrt{1-4 r_\ell^2/s}$, $\lambda(a,b,c)=a^2+b^2+c^2-2(ab+ac+bc)$, $s=q^2/m_h^2$,  and $r_i=m_i/m_h$. Here $\theta^{(\ell)}$ is the angle between the three-momentum of Higgs boson and the three-momentum of $\ell^-$ in the dilepton rest frame, and the phase space is given by
\beq\label{phasespace}
4 r_\ell^2\leq s \leq (1-r_V)^2,\;\;\;\; -1\leq\cos\theta^{(\ell)}\leq 1.
\eeq

 Let us further analyze $h\to\pi^0\bar{\ell}\ell$ modes with $\ell$ denoting charge leptons or neutrinos. In the SM, these processes will get dominant contributions through $hZZ$ vertex at the tree level, and their amplitude can be directly written as
\beq\label{amplitude-hpill}
{\cal M}(h\to\pi^0\bar{\ell}\ell)=-\frac{g^2 m_Z^2 f_\pi p^\mu}{2\sqrt{2}v \cos^2\theta_W(m_Z^2-m_\pi^2)}P_Z \bar{u}_\ell(k_1) \gamma_\mu(g_V^\ell-g_A^\ell\gamma_5)v_\ell(k_2),
\eeq
where $p$ is the momentum of the neutral pion. Similar diagrams like Figure 1(b), in which $\pi^0$ is converted through the virtual $Z$, can also lead to the transition involving charged lepton final states at the tree level. However, it is easy to see that these diagrams will be suppressed by $m_\ell^2/m_Z^2$, comparing with eq. (\ref{amplitude-hpill}), which thus could be negligible.

To be complete, we also check the lowest order dominant contribution to $h\to P^+ \ell^-\bar{\nu}_\ell$ ($P=\pi, K$) and $h\to V^+\ell^-\bar{\nu}_\ell$ ($V=\rho, K^{*}$) decays, which are generated by the Higgs coupling to $W$-boson ($h W^+ W^-$) vertex in the SM. As mentioned above, now we also neglect those strongly suppressed diagrams like Fig. 1(b), in which $Z$-boson is replaced by $W$-boson.  Thus their decay amplitudes can be easily found to be
\beq\label{amplitude-hpnul}
{\cal M}(h\to P^+ \ell^-\bar{\nu}_\ell)=\frac{g^2 m_W^2 f_P V_{ij}}{4 v (m_W^2-m_P^2)}P_W p^\mu \bar{u}_\ell(k_1) \gamma_\mu(1-\gamma_5)v_\nu(k_2),
\eeq
 and
 \beq\label{amplitude-hrhonul}
{\cal M}(h\to V^+ \ell^-\bar{\nu}_\ell)=\frac{g^2 m_W^2 f_V m_V V_{ij}}{4 v (m_W^2-m_V^2)}P_W \epsilon^{*\mu}_V(p) \bar{u}_\ell(k_1) \gamma_\mu(1-\gamma_5)v_\nu(k_2)
\eeq
with
\beq\label{PW}
P_W=\frac{1}{m_W^2-q^2-im_W\Gamma_W}
\eeq
parameterizing the $W$ pole effect. Here $V_{ij}$ denotes the relevant CKM matrix element, which is equal to $V_{ud}$ for $\pi^+$ and $\rho^+$, and $V_{us}$ for $K^+$ and $K^{*+}$, respectively.

Similar to the case of $h\to V\ell^+\ell^-$ decays, as shown in eq. (\ref{distribution1}), it is easy to derive the differential decay rates of the above other processes.

\section{Numerical analysis}

%\subsection{Theoretical branching ratios}

To illustrate the numerical results for branching fractions of exclusive Higgs decays to lepton pair plus light mesons, we normalize these decay rate to the theoretical prediction for the total Higgs width in the SM, $\Gamma_h=4.10$ MeV, referring to $m_h=125.09$ GeV \cite{LHCHiggsgroup}. Thus for $h\to V \bar{\ell}\ell$ with $V=\rho^0$ and $\omega$, it is straightforward to obtain
\beqn\label{br1}
{\cal B}(h\to\rho^0 e^+e^-)=(4.61\pm 0.06)\times 10^{-7},\\
{\cal B}(h\to\rho^0\mu^+\mu^-)=(3.97\pm 0.05)\times 10^{-7},\\
{\cal B}(h\to\rho^0\tau^+\tau^-)=(1.80\pm 0.02)\times 10^{-5},\\
{\cal B}(h\to\omega e^+e^-)=(4.17\pm 0.09)\times 10^{-8},\\
{\cal B}(h\to \omega \mu^+\mu^-)=(3.61\pm 0.08)\times 10^{-8},\\
{\cal B}(h\to\omega\tau^+\tau^-)=(1.58\pm 0.03)\times 10^{-6}
\eeqn
for charge lepton final states, and for neutrino final states
\beqn
{\cal B}(h\to\rho^0 \nu\bar{\nu})=(1.19\pm 0.01)\times 10^{-6},\\\label{br2}
{\cal B}(h\to\omega\nu\bar{\nu})=(1.11\pm 0.02)\times 10^{-7},
\eeqn
where a factor 3 has been included in the calculation due to the neutrino flavors.

Furthermore, by taking the values of decay constants from Ref. \cite{KN15}, as shown in Table 1, we get
\beqn\label{br3}
{\cal B}(h\to\pi^0 e^+e^-)=(7.06\pm 0.02)\times 10^{-8},\\
{\cal B}(h\to\pi^0 \mu^+\mu^-)=(7.06\pm 0.02)\times 10^{-8},\\
{\cal B}(h\to\pi^0 \tau^+\tau^-)=(7.10\pm 0.02)\times 10^{-8},\\
{\cal B}(h\to\pi^0 \nu\bar{\nu})=(4.21\pm 0.01)\times 10^{-7},
\eeqn
and
\beqn\label{br4}
{\cal B}(h\to\pi^+ \ell^-\bar{\nu}_\ell)=(4.05\pm 0.01)\times 10^{-7},\\
{\cal B}(h\to\rho^+ \ell^-\bar{\nu}_\ell)=(1.03\pm 0.01)\times 10^{-6},\\
{\cal B}(h\to K^+ \ell^-\bar{\nu}_\ell)=(3.14\pm 0.01)\times 10^{-8},\\
{\cal B}(h\to K^{*+} \ell^-\bar{\nu}_\ell)=(5.32\pm 0.02)\times 10^{-8}
\eeqn
for $\ell=e,\mu$, and $\tau$, respectively. The errors of our predictions are due to the uncertainties of the decay constants of light mesons listed in Table 1.

 Note that our predicted branching ratios are in the range of $10^{-5}\sim 10^{-8}$, which can be compared to those predicted for similar processes involving heavy meson by the authors of Ref. \cite{CFS16}: ${\cal B}(h\to J/\psi\mu^+\mu^-)=(9.10\pm 0.50)\times 10^{-8}$, ${\cal B}(h\to J/\psi \tau^+\tau^-)=(1.82\pm 0.10)\times 10^{-6}$, ${\cal B}(h\to \Upsilon\mu^+\mu^-)=(5.60\pm 0.37)\times 10^{-8}$, and ${\cal B}(h\to \Upsilon\tau^+\tau^-)=(5.66\pm 0.29)\times 10^{-8}$. Experimentally, it is certainly challenging to search for these rare decays due to the smallness of their decays rates. On the other hand, as mentioned in the introduction section, for the final state involving light mesons, for instance in $h\to \rho \ell\bar{\ell}$ transitions $\rho$ meson is reconstructed through $\rho\to \pi\pi$ with very large branching ratio; while this is not the case for heavy quarkonia final states.

 One can find that, for different charged lepton flavors ($e$,$\mu$, and $\tau$), the decay rates of $h\to P^+(V^+) \ell^-\bar{\nu}_\ell$ are degenerate, and those of $h\to \pi^0\ell^+\ell^-$ are almost degenerate. This is easily understood because the former is dominated by the virtual $h\to W^+W^-$ transition and the latter is by the virtual $h\to Z Z$ vertex.

However, it is a different story for $h\to V\ell^+\ell^-$ processes. The modes with $\tau$ pairs $h\to \rho^0(\omega) \tau^+\tau^-$ are predicted with larger rate while the channels with muons and electrons have rates suppressed about by a factor 40, and
 \beq\label{ratioemu}
 \frac{{\cal B}(h\to \rho^0\mu^+\mu^-)}{{\cal B}(h\to \rho^0 e^+ e^-)}\simeq\frac{{\cal B}(h\to \omega\mu^+\mu^-)}{{\cal B}(h\to \omega e^+ e^-)}\simeq 0.86.
 \eeq
 In the present case, as shown in Figure 1,  all of the $ZZ$, $Z\gamma$, $\gamma\gamma$ and $\ell^+ \ell^-$ intermediate states can give contributions.

\begin{figure}[t]
\begin{center}
\includegraphics[width=5.2cm,height=3.6cm]{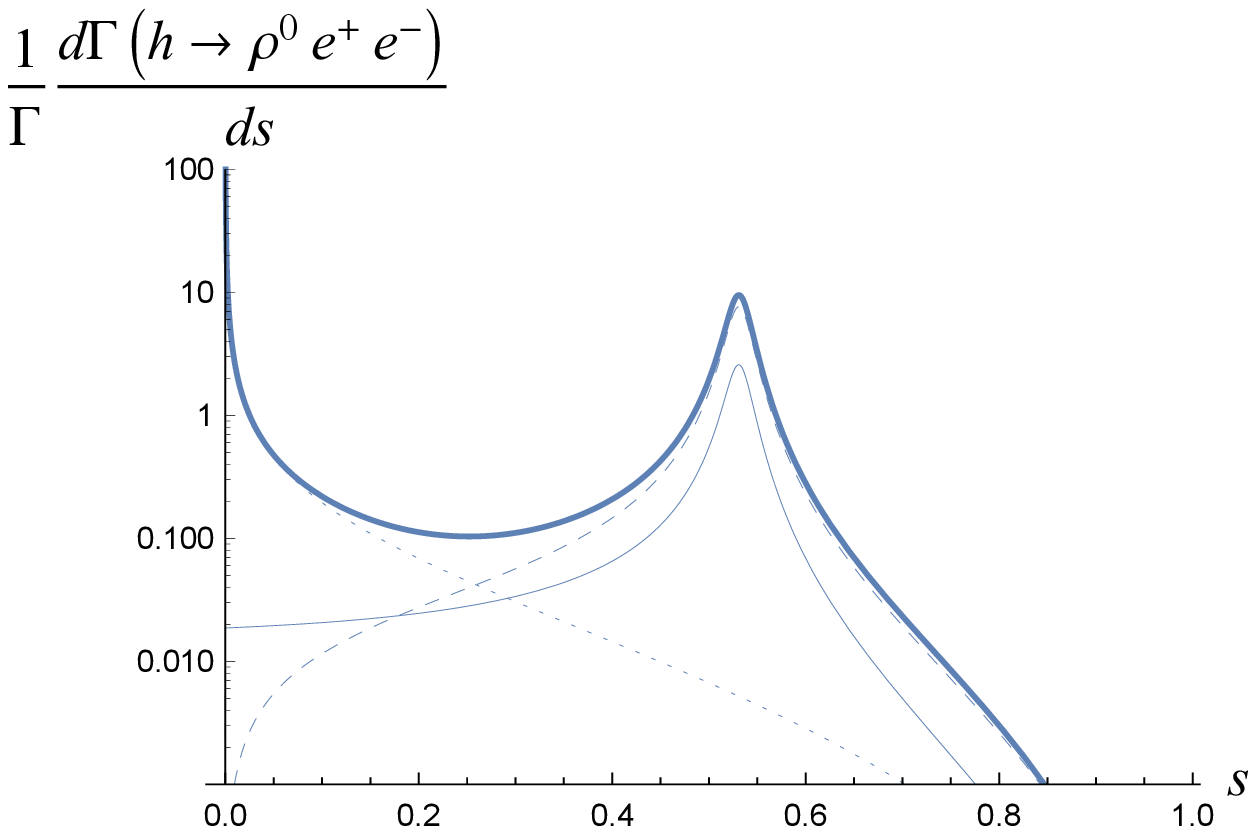}
\includegraphics[width=5.2cm,height=3.6cm]{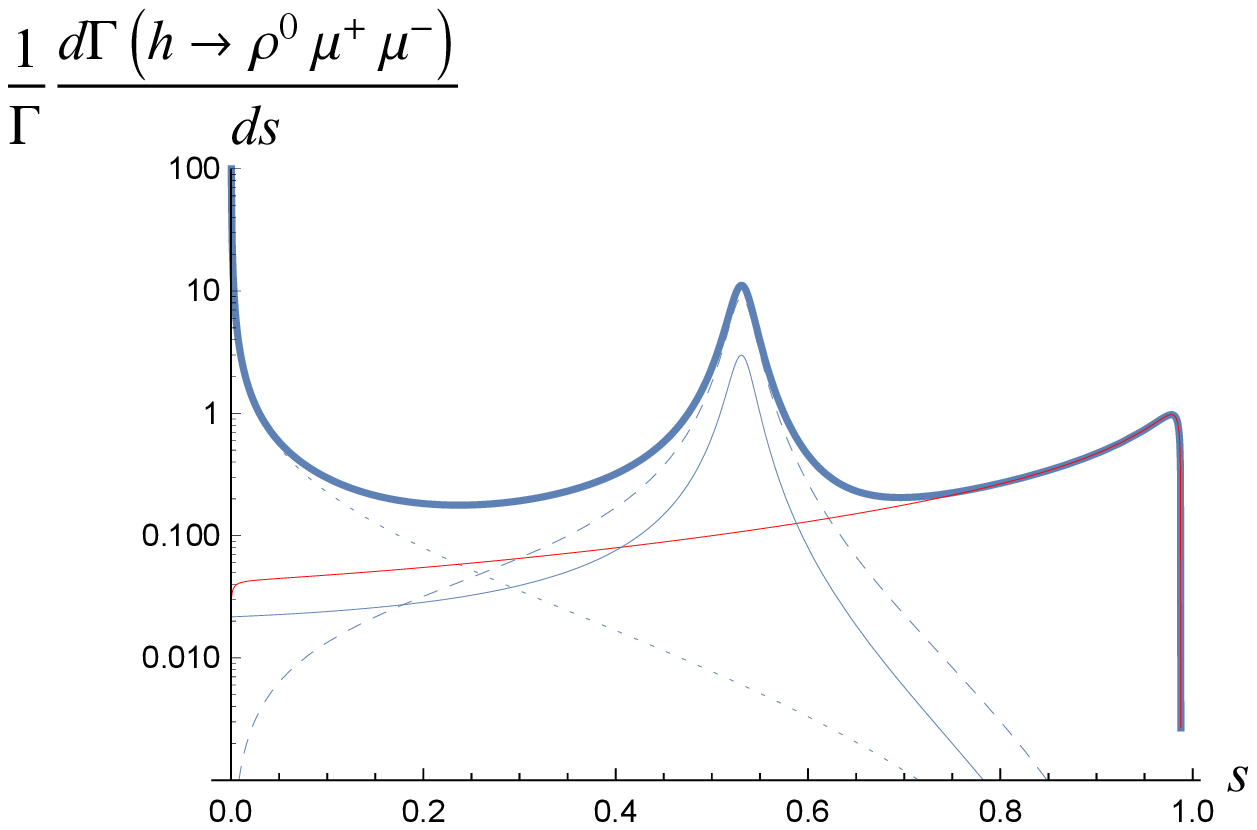}
\includegraphics[width=5.2cm,height=3.6cm]{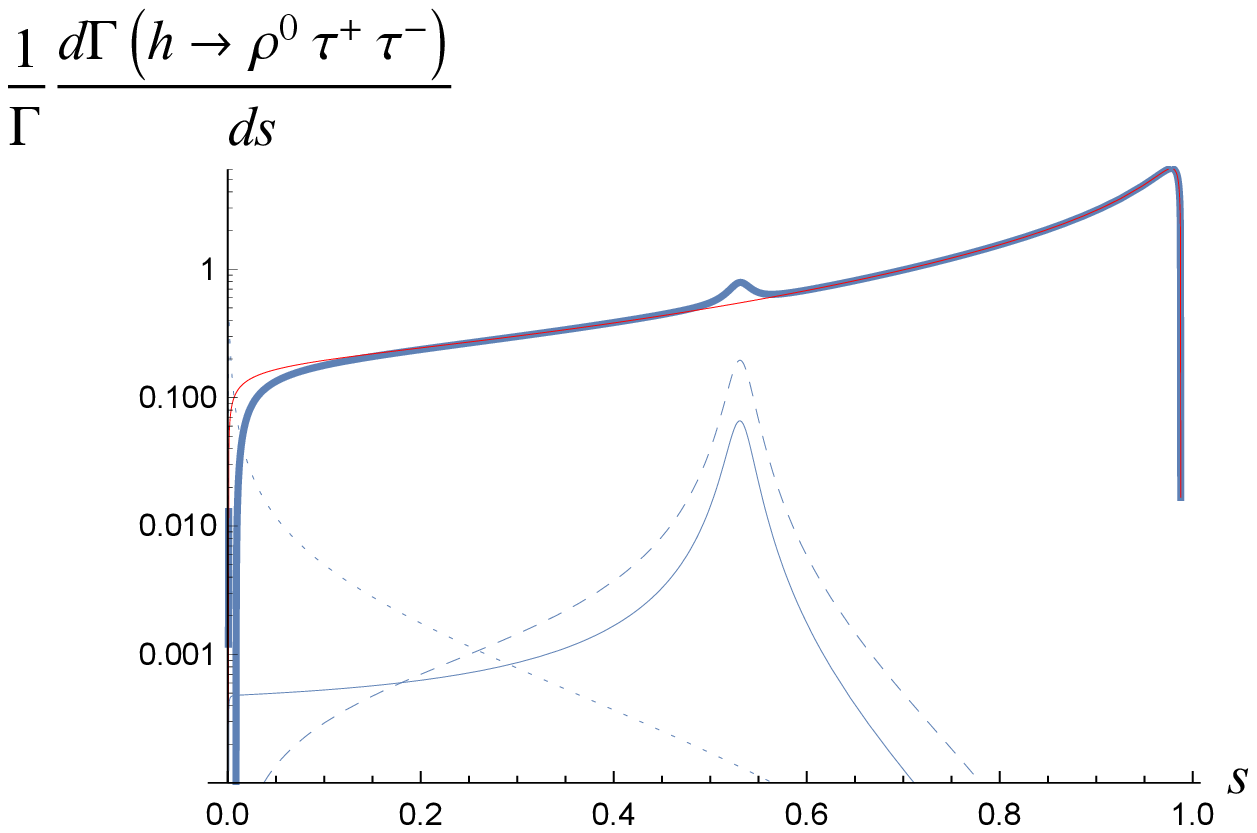}
\end{center}\caption{The normalized invariant mass distribution of $h\to\rho^0 \ell^+\ell^-$ decay for $\ell=e,\mu$, and $\tau$, respectively, with $s=q^2/m_h^2$ and $q^2$ is dilepton mass squared. The thin solid line denotes the contribution from $hZZ$, the dashed line denotes the contribution from $hZ\gamma$, and the dotted line the contribution from $h\gamma\gamma$, and the red line the contribution from Figure 1(b) via the virtual photon only while the thick solid line gives the total contribution. For the electron mode, the contribution from Figure 1(b) is vanishingly small, which cannot be shown in the plot.}\label{figure2}
\end{figure}

 In order to explicitly observe their roles for different charged lepton final states, let us take $h\to\rho^0\ell^+\ell^-$ as examples, and the dilepton invariant mass distributions of these decays have been displayed in Figure \ref{figure2} for $\ell=e,\mu,$ and $\tau$, respectively. Different types of contributions are plotted separately for comparison. From these three plots, one can readily find that, the contribution from the Higgs coupling to leptons [Figure 1(b)] is vanishingly small for the electron mode, and could be relevant in the muon case; while it is dominant, giving about 99\% contribution, thus leads to larger rate for the $\tau^+\tau^-$ final state. Other contributions are strongly suppressed in the process $h\to \rho^0\tau^+\tau^-$.

 On the contrary, for $\ell=e$ and $\mu$, both of them are dominated by $Z\gamma$, $\gamma\gamma$, and $ZZ$ intermediate states via the $\gamma^*/Z$ poles, corresponding to the two peaks in above plots. Obviously, $\gamma\gamma$ plays the dominant role in the low dilepton mass region. When increasing the dilepton mass, other contributions from $Z\gamma$ and $ZZ$ will become dominant but $Z\gamma$ gives the more important one. At the large dilepton mass region after the $Z$ pole, the contribution from $\ell^+\ell^-$ intermediate state [Figure 1(b) via the virtual photon] will be relevant for the muon mode. Further detailed numerical analysis shows that $hZZ$ and $hZ\gamma$ give almost the same contributions to both modes. However, $h\gamma\gamma$ and the Higgs coupling to leptons play different roles in these two processes. The electron mode can obtain the relative large contribution from $h\gamma\gamma$ in the very low dilepton mass region due to the smallness of the electron mass, and as discussed already, a significant contribution from Figure 1(b) at the high dilepton mass region could be expected in $h\to\rho^0\mu^+\mu^-$ decay, which thus leads to the ratio in eq. (\ref{ratioemu}). In addition, it is interesting to note that, in the SM, the $hZ\gamma$ and $h\gamma\gamma$ couplings are loop-induced while $hZZ$ and the Higgs coupling to leptons are the tree-level vertices. This seems to indicate that $h\to \rho^0\ell^+\ell^-$ with $\ell=e$ and $\mu$ could be sensitive to the short-distance physics and studies of these decays may help to probe the novel dynamics in the Higgs sector.

\begin{figure}[t]
\begin{center}
\includegraphics[width=5.2cm,height=3.6cm]{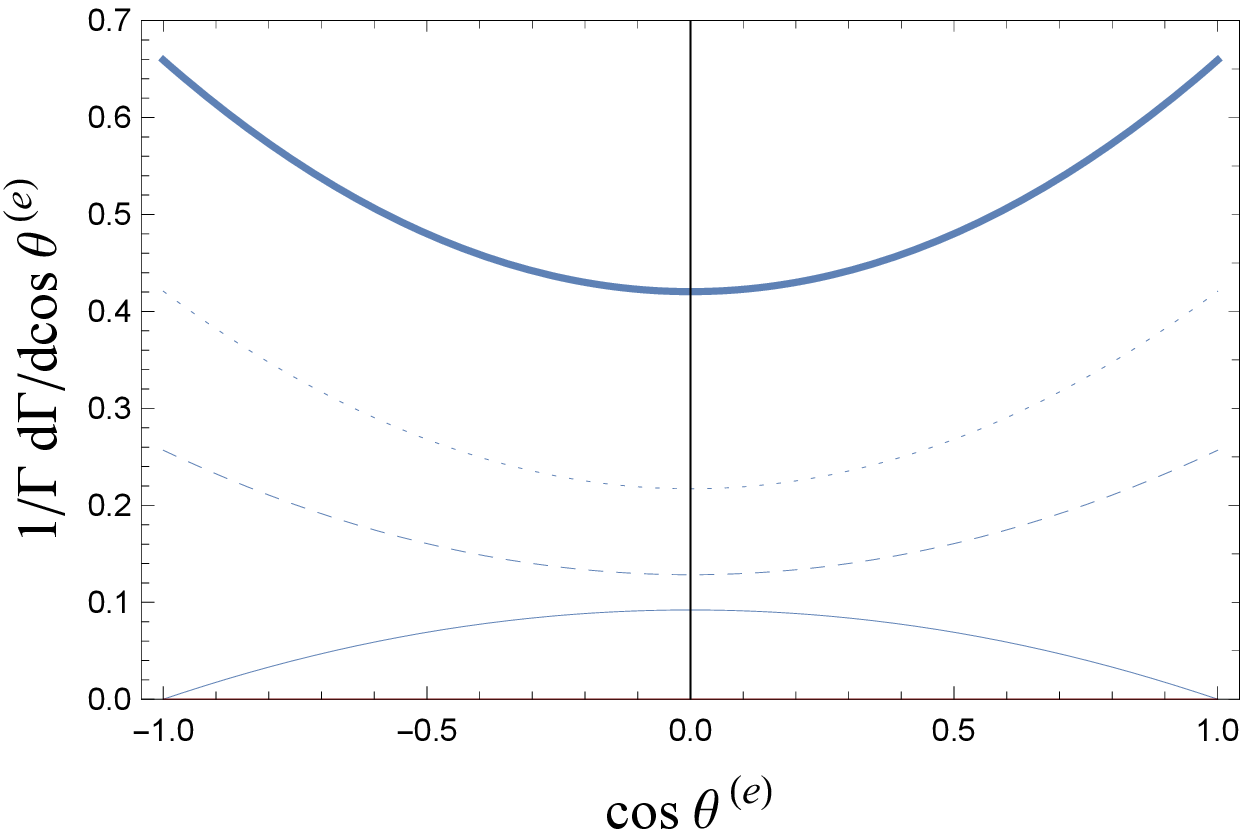}
\includegraphics[width=5.2cm,height=3.6cm]{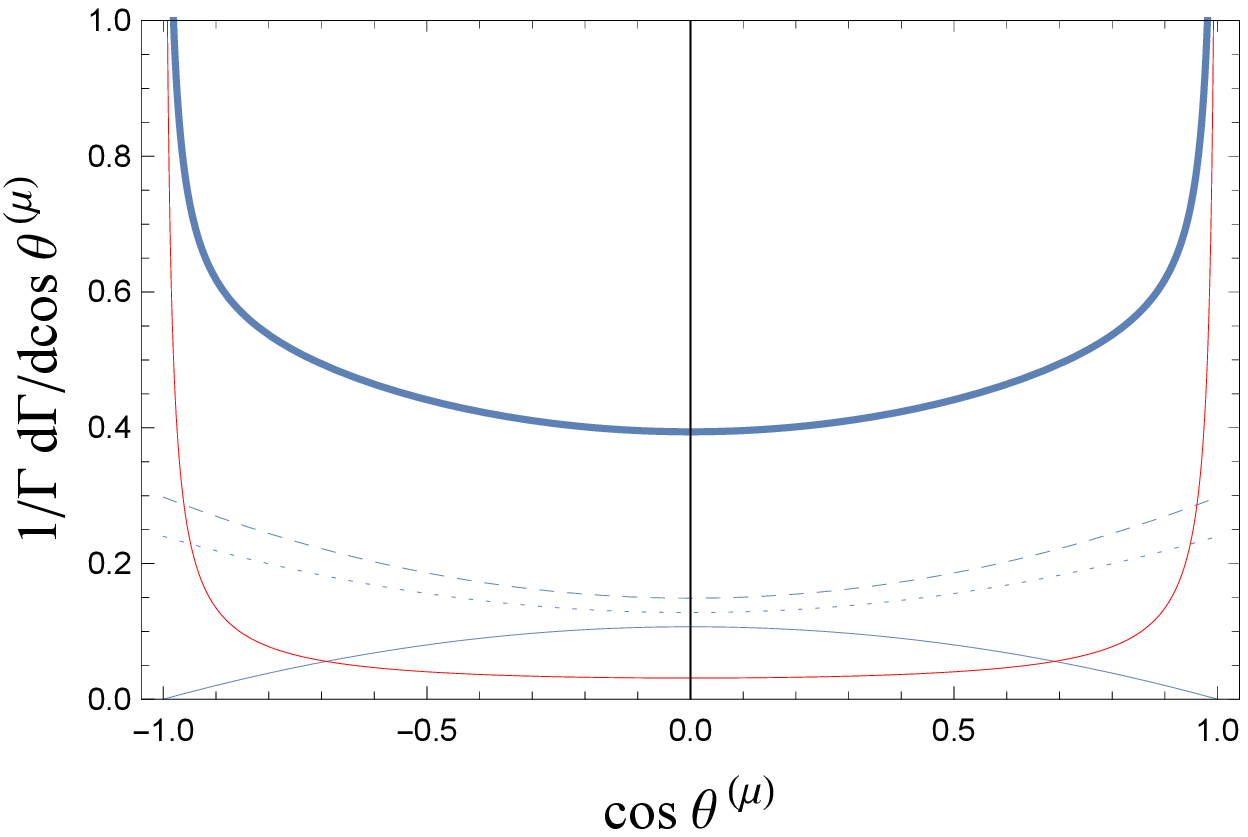}
\includegraphics[width=5.2cm,height=3.6cm]{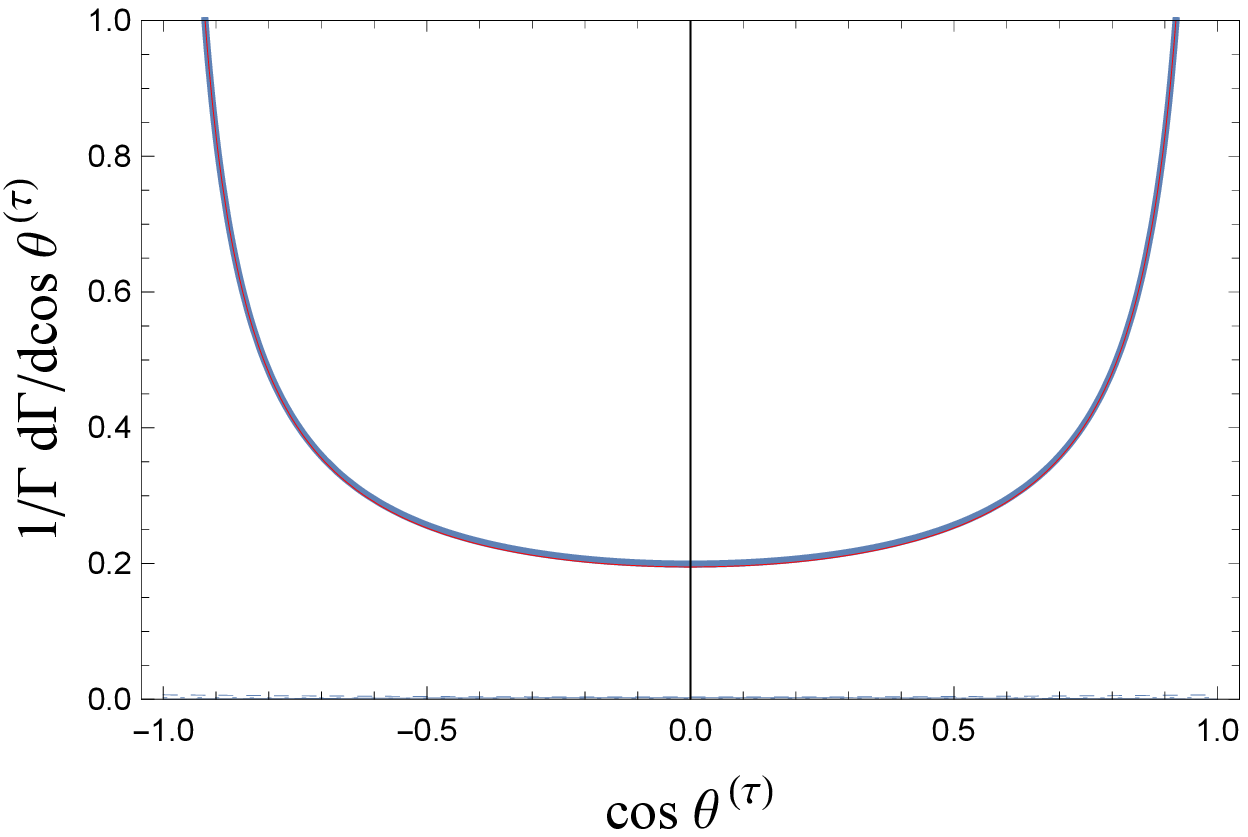}
\end{center}
\caption{The normalized angular distribution of  $h\to\rho^0 \ell^+\ell^-$ with respect to $\cos\theta^{(\ell)}$ for $\ell=e,\mu$, and $\tau$, respectively. The thin solid line denotes the contribution from $hZZ$, the dashed line denotes the contribution from $hZ\gamma$, the dotted line the contribution from $h\gamma\gamma$, and the red line the contribution from Figure 1(b) via the virtual photon only while the thick solid line gives the total contribution.}\label{figure3}
\end{figure}

 It is noticed that, besides the dilepton invariant mass distributions, one can further examine the differential angular distributions of $h\to\rho^0\ell^+\ell^-$ decays by integrating over $s$ in eq. (\ref{distribution1}). The resulting distributions with respect to $\cos\theta^{(\ell)}$ are given in Figure \ref{figure3} for $\ell=e,\mu$, and $\tau$, respectively. Similar to the case of the invariant mass distribution, the contribution from Figure 1(b) almost saturates the angular differential decay rate in $h\to\rho^0\tau^+\tau^-$, can be negligible in $h\to\rho^0 e^+e^-$, and is not very large but non-negligible for the muon mode. It is easy to see that, the angular distribution is symmetric for $\cos\theta^{(\ell)}\leftrightarrow -\cos\theta^{(\ell)}$, therefore no forward-backward asymmetry could be expected in $h\to\rho^0\ell^+\ell^-$ if considering these transitions only in the SM.

\begin{figure}[t]
\begin{center}
\includegraphics[width=7cm,height=5cm]{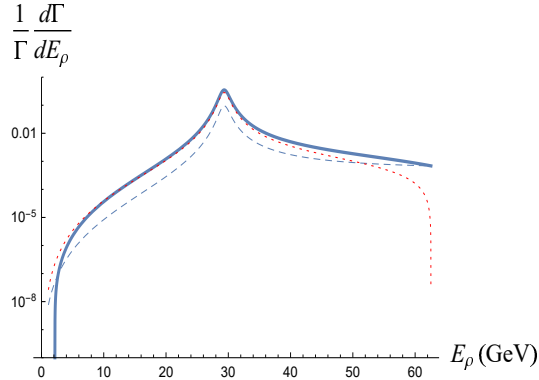}
\end{center}
\caption{The normalized energy spectrum of  $h\to\rho^0 \nu \bar{\nu}$ with respect to $E_\rho$. The dashed line denotes the contribution from $hZZ$, the red-dotted line the contribution from $hZ\gamma$, and the thick solid line gives the total contribution.}\label{figure4}
\end{figure}

We also study the energy spectrum of the processes involving neutrino final states. Now the lepton pair mass squared $q^2=m_h^2-2 m_h E_\rho+m_\rho^2$, where $E_\rho$ denotes the energy of $\rho$ meson in the rest frame of Higgs boson. Using eq. (\ref{phasespace}), we have $m_\rho\leq E_\rho\leq (m_h^2+m_\rho^2)/2 m_h$. Thus the normalized energy spectrum of $h\to\rho^0\nu\bar{\nu}$ decay with respect to $E_\rho$ can be plotted, which is displayed in Figure \ref{figure4}. As mentioned above, this channel gets the dominant contribution from $hZZ$ and $hZ\gamma$ transitions. It is shown that the plot has a peak when $E_\rho\sim 30$ GeV. The reason for this is that $\sqrt{q^2}$ is close to $m_Z$ and the virtual $Z$ boson can be on-shell around this region.

From eqs. (\ref{br1}) $-$ (\ref{br2}), it is found that decay rates of $h\to \omega \ell^+\ell^-$ and $h\to \omega \nu\bar{\nu}$ are suppressed about one order of magnitude, compared to those of $\rho^0$ meson's. However, it is believed that one will achieve the similar behaviors as above if we also perform the dilepton invariant mass and angular distribution analysis of these decays.

\section{Summary and outlook}

We have investigated exclusive rare Higgs decays including $h\to \rho^0(\omega)\ell\bar{\ell}$, $h\to \pi^0\ell\bar{\ell}$, $h\to \pi^+ (K^+)\ell^-\bar{\nu}_\ell$, and $h\to \rho^+(K^{*+})\ell^-\bar{\nu}_\ell$ in the SM.  Decay rates of these modes have been calculated, and their branching ratios are predicted around $10^{-5}\sim 10^{-8}$, which can be compared to $h\to J/\psi (\Upsilon)\ell\bar{\ell}$ decays. Experimental observations of these rare processes are generally challenging, which, however, might be interesting in the future high energy and high-precision experiments, in order both to test the SM and to search for new physics (NP) beyond the SM.

We have presented a detailed analysis of the differential dilepton invariant mass distributions of $h\to \rho^0 \ell^+\ell^-$ decays. In the SM, decay amplitudes of these channels are governed by Higgs couplings to gauge bosons ($hZZ$, $hZ\gamma$, and $h\gamma\gamma$ vertices) and couplings to leptons ($h\ell\bar{\ell}$ vertex). It has been shown that the contribution from the $h\ell\bar{\ell}$ vertex is completely dominant in $h\to \rho^0 \tau^+\tau^-$ and negligible in $h\to \rho^0 e^+ e^-$. But all types of vertices can play significant roles in $h\to \rho^0\mu^+\mu^-$. In the SM, the $hZZ$ and $h\ell\bar{\ell}$ couplings exist at the tree level while the $h\gamma\gamma$ and $hZ\gamma$ vertices are loop-induced and hence suppressed. However, due to the smallness of the light vector meson mass, the photon propagator will be almost on-shell, which thus counteracts the loop suppression. As a result, the $hZZ$, $h\gamma\gamma$, and $hZ\gamma$ couplings are of similar importance for the $e^+e^-$ and $\mu^+\mu^-$ final states. Particularly, the contribution from the $hZ\gamma$ vertex even dominates over that from the $hZZ$ interaction in the large dilepton mass region, as displayed in the plots of Figure \ref{figure2}. Thus the deviation from the SM prediction may be observed in $h\to \rho^0\ell^+\ell^-$ if NP scenarios can give rise to significant enhancement to the $hZ\gamma$ coupling.

From recent measurements by ATLAS and CMS Collaborations \cite{cmsatlas19}, it is known that the Higgs couplings to the SM gauge bosons and fermions agree with the SM predictions within experimental and theoretical uncertainties. In particular, Higgs boson decays like $h\to ZZ^*$ \cite{cms19-zz}, $h\to \gamma\gamma$ \cite{atlas18-gg}, and $h\to \mu^+\mu^-$ \cite{atlascms20-mumu} have been well studied and their rates are consistent with the SM expectations at or below the ${\cal O}(20\%)$ level. However, the channel $h\to Z \gamma$ has not been detected yet. The most recent search for this mode at the LHC comes from ATLAS Collaboration, and the current strongest upper bound on its decay rate is set at 3.6 times the SM value \cite{atlas20-zgamma}. This means that the present experimental limit allows substantial room for NP in $h\to Z\gamma$ decay.

Some interesting NP models have recently been constructed \cite{ASV20}, in which the large contributions to $h\to Z\gamma$ can be generated, close to the current experimental upper bound, without in conflict with all other Higgs measurements. Accordingly, significant enhancement to decay rates of $h\to\rho^0\ell\bar{\ell}$ could be expected. However, as pointed out by the authors of Ref. \cite{ASV20}, this can only occur in some complicated NP sectors. In general, it is impossible to achieve the above goal in simple models.
Thus it is reasonable to anticipate that, with the accumulation of more experimental data, the decay rate of $h\to Z\gamma$ would be observed close to the SM expectation at or below ${\cal O}(20\%)$ level eventually, like that of other Higgs decays.

On the other hand, even in the case that the NP effects are around the ten percent level, three-body processes $h\to\rho^0\ell^+\ell^-$ ($\ell=e, \mu$), or experimentally four-body $h\to (\pi^+\pi^-)_\rho \ell^+\ell^-$,  may offer nontrivial invariant mass and angular distributions, which would provide interesting information on short-distance dynamics. Therefore, if new structures other than eqs. (\ref{effectivehzgamma}) and (\ref{effectivehgammagamma}), for instance, the parity-odd term like $\epsilon^{\mu\nu\alpha\beta} h Z_{\mu\nu}F_{\alpha\beta}$, appear in NP scenarios, some useful observables such as forward-backward asymmetries or other asymmetries, generated from the interference between new amplitudes and the SM ones, will come out in the angular analysis.  This is similar to angular studies of $h\to Z\ell^+\ell^-$ decays in Refs. \cite{BCD13, BBW14}. Such detailed analysis of the $h\to (\pi^+\pi^-)_\rho \ell^+\ell^-$ angular distribution is beyond the scope of the present paper, which is left for a future separate publication.

\section*{Acknowledgements}
This work was supported in part by the National Natural Science Foundation of China under Grants No. 11575175, No. 11947301, and No. 12047502, and by National Key Basic Research Program of China under Contract No. 2020YFA0406400.

\end{document}